\documentclass[a4paper,12pt]{article}
\usepackage{graphicx}
\usepackage{amssymb,latexsym}
\usepackage[english]{babel}
\usepackage{graphicx}
\newcommand{\bec}{\begin{center}}
\newcommand{\ec}{\end{center}}
\newcommand{\bee}{\begin{equation}}
\newcommand{\ee}{\end{equation}}

\begin{document}
\large
\begin{titlepage}
\begin{center} 
{\Large\bf {Decay width modeling of Higgs boson within THDM model}}\\
\vspace*{12mm} {\bf  T.V. Obikhod and I.A. Petrenko \\}
\vspace*{10mm}
{\it Institute for Nuclear Research\\
National Academy of
Sciences of Ukraine\\
03068 Kiev, Ukraine\\}
e-mail: obikhod@kinr.kiev.ua\\
\vspace*{21mm}
{\bf Abstract\\}
\end{center}
As part of the search for new physics beyond the Standard Model, we chose the determination of the Higgs boson decay width as one of the least experimentally determined values. The decay widths into the four fermions of the lightest and heaviest CP-even Higgs bosons of the THDM model were calculated, taking into account QCD and electroweak corrections in the NLO approximation. To achieve this goal, the program Monte Carlo Prophecy 4f with special scenarios of parameters, 7B1 and 5B1 were used. It was found that the decay width of the heavier CP-even Higgs boson, H differs from H$_{SM}$ by 1227.93 times and changes to a negative value when deviating from the standard scenarios. Scale factors k$^2_{Z}$ and k$^2_{W}$ showed the predominance of the associated with Z boson production cross section of CP-even Higgs boson over the associated with W production cross section.
\vspace*{8mm}\\

\end{titlepage}

\begin{center}
\textbf{\textsc{1. Introduction}}
\end{center}
	In light of the latest experimental data on the searches for new physics beyond the Standard model (SM), Higgs boson remains the only candidate for a window into new physics, \cite{1.}. This task is related to the experimental study and theoretical predictions of the properties of the Higgs boson: the production cross sections, the partial decay width, coupling measurements, $k_i$. The crucial role for the investigation of the Higgs boson properties is played by the Higgs branching ratios and decay widths, \cite{2.}.
	The Higgs particle is a massive scalar boson with zero spin, no electric charge, and no colour charge is very unstable, decaying immediately into other particles. As all the channels of decay of the Higgs boson as well as possible new particles with certain masses have not yet been studied, there are uncertainties in the properties of the coupling constants and, accordingly, in the decay width of this particle. This fact is demonstrated by the deviation of the predicted SM Higgs decay width of about $4.07 \cdot 10^{-3}$ GeV from the experimental data, which are presented in Table 1, \cite{3., 4.}.
\vspace{3 mm}
\bec	
\emph{\textbf{\small Table 1.}} {\it\small  Run 1 observed (expected) direct 95\% CL constraints on the width of the
125 GeV resonance from fits to the $\gamma\gamma$ and ZZ mass spectra. The CMS measurement
from the 4l mass line-shape was performed using Run 2 data.}
\ec	
\bec	
\begin{tabular}{lccl}\hline
Experiment                        & $M_{\gamma\gamma}$  & $M_{4l}$ &  \\ 
\hline
ATLAS 	 & $\prec 5.0(6.2) $ GeV      & $\prec 2.6(6.2) $   GeV    &  \\
CMS      & $\prec 2.4(3.1) $   GeV    & $\prec 1.1 (1.6) $   GeV   &  \\  
\end{tabular}
\ec
\vspace{0 mm}

	The purpose of our paper is to calculate decay widths of lightest, h, and CP-even, H, Higgs bosons of Two Higgs doublet model (THDM), \cite{5.} as well as the value of the deviation from SM of the sum of the partial Higgs decay widths compared to the SM, $\kappa_H^2$, through computer modeling with the help of Monte Carlo program Prophecy 4f 3.0 \cite{6.}. 
\begin{center}
\textbf{\textsc{2. The calculations of decay width and scale factors}}
\end{center}
	The Standard Model predicts a very small width of about $4$ MeV for a $126$ GeV Higgs boson. But the error of the energy measurement at the LHC is hundreds of times greater, of the order of $1$ GeV, and it will not be possible to significantly reduce it. As a result, measuring the width of the Higgs boson directly is unrealistic. However, it is possible to accumulate data on the production and decay of the Higgs boson at significantly higher energies - not in the vicinity of $126$ GeV, but, say, above $300$ GeV, \cite{7.}. This process will look like the birth and decay of a virtual Higgs boson in this mass range. It is, of course, strongly weakened in comparison with the main process at the resonance peak, but it can be quite measurable.
	
	As THDM model predicts the existence of five Higgs bosons, we will carry out our calculations for two bosons: lightest Higgs boson, h and CP-even Higgs boson, H as the analog of virtual Higgs boson described above. Thus, the idea of theorists - to accumulate data on the production and decay of the Higgs boson at significantly higher energies can be realized. The efficiency of this method can be estimated by comparing the calculations of decay widths for the lightest and heaviest bosons. 
	
	The precise experimental investigation of the Higgs boson and theoretical searches for deviations from the predictions of $SM$ requires precise Monte Carlo computer modeling. Prophecy 4f computes the inclusive partial decay widths and differential distributions of the decay products, where unweighted events for leptonic final states are provided. The advantage of the Prophecy4f program is that it allows the calculations for the Higgs decays into four fermions including full electroweak and QCD next-to-leading order (NLO) corrections with interference contributions between different $WW$/$ZZ$ channels, and inclusion of all off-shell effects of intermediate $W/Z$ bosons.
	
	We'll consider the processes, LO Feynman diagram of which is in the form of Fig. 1 
\vspace{1 mm}
\bec
\includegraphics[width=0.36\textwidth]{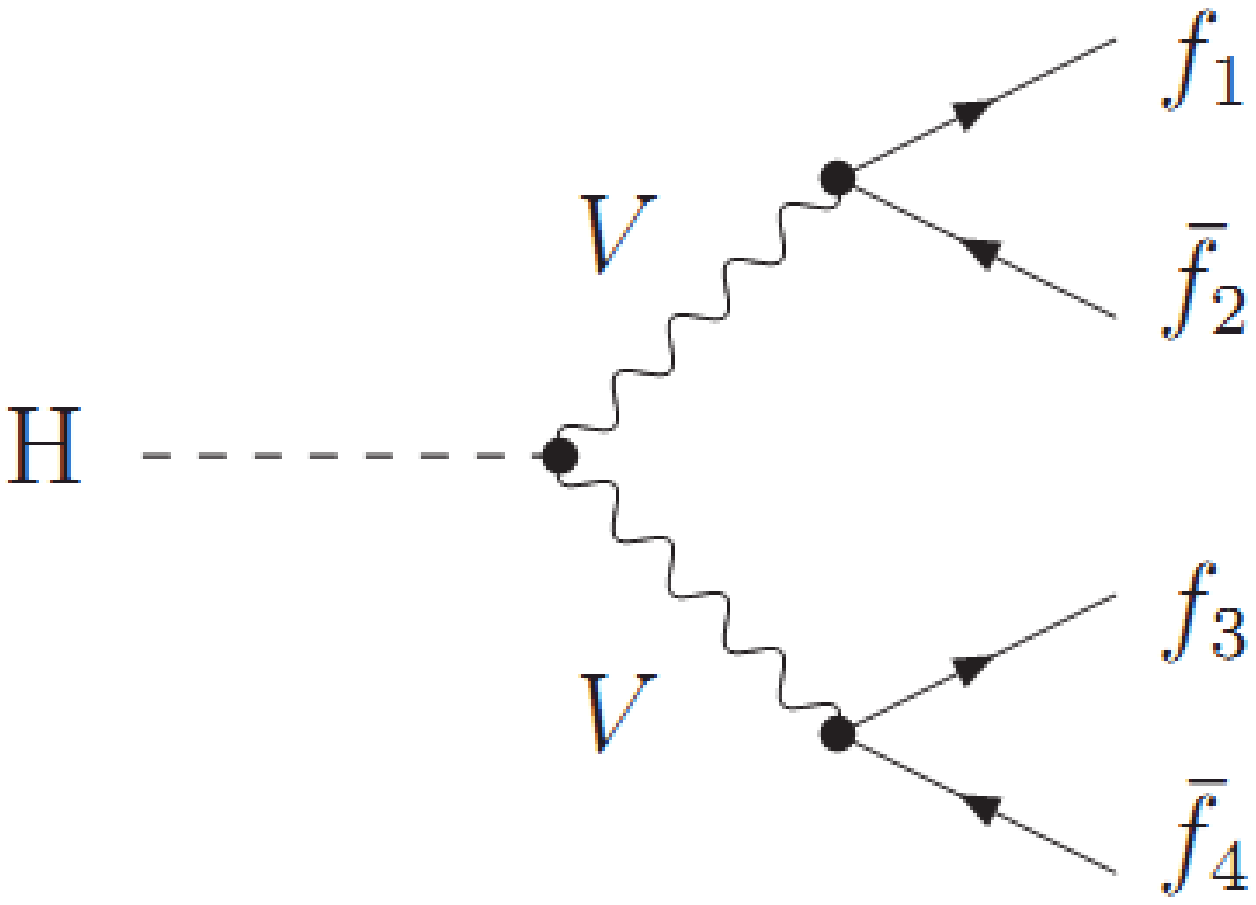}\\
\vspace{5 mm}
\emph{\textbf{Fig.1.}} {\emph{ Generic diagram for decay of $H \rightarrow 4f$ where $V = W, Z$, from \cite{6.}. }}\\
\ec
\vspace*{1mm}
	The total state width of Higgs boson is equal to the sum of the partial channel widths \cite{6.}:

\[\Gamma_{H\rightarrow 4f} = \Gamma^{{total}} = \Gamma^{{leptonic}} + \Gamma^{{semi-leptonic}} + \Gamma^{{hadronic}},
	\]

	The total width can be presented via $ZZ$, $WW$ decays and their interference:
	\[
\Gamma_{H\rightarrow 4f} =\Gamma_{H\rightarrow W^*W^*\rightarrow 4f} +\\
+ \Gamma_{H\rightarrow Z^*Z^*\rightarrow 4f} + \Gamma_{WW/ZZ-int}\ ,
	\]
where the components are defined in terms of specific final states: 
	\\
	\\
$\Gamma_{H\rightarrow W^{\star}W^{\star}\rightarrow 4f} 
=9\cdot \Gamma_{H\rightarrow \nu_e \overline{e} \mu^{-}\nu_{\mu}} + 12\cdot \Gamma_{H\rightarrow \nu_e \overline{e}d\overline{u}} + 4\cdot \Gamma_{H\rightarrow u \overline{d}s\overline{c}} \ ,
$
\\
\\
$
\Gamma_{H\rightarrow Z^*Z^*\rightarrow 4f} = 3\cdot \Gamma_{H\rightarrow  \nu_e \overline{\nu_e}\nu_{\mu} \overline{\nu_{\mu}}} + 3\cdot \Gamma_{H\rightarrow e \overline{e}\mu \mu^+} \\ 
\hspace*{2.6cm}+ 9\cdot \Gamma_{H\rightarrow \nu_e \overline{\nu_e}\mu \mu^+} + 3\cdot \Gamma_{H\rightarrow \nu_e \overline{\nu_e}\nu_e \overline{\nu_e}} \\
\hspace*{2.6cm}+ 3\cdot \Gamma_{H\rightarrow e \overline{e}e \overline{e}} + 6\cdot \Gamma_{H\rightarrow \nu_e \overline{\nu_e}u \overline{u}} \\
\hspace*{2.6cm}+ 9\cdot \Gamma_{H\rightarrow \nu_e \overline{\nu_e}d \overline{d}} + 6\cdot \Gamma_{H\rightarrow u\overline{u}e \overline{e}}\\
\hspace*{2.6cm}+ 9\cdot \Gamma_{H\rightarrow d \overline{d}e\overline{e}} + 1\cdot \Gamma_{H\rightarrow u\overline{u}c\overline{c}}\\
\hspace*{2.6cm}+ 3\cdot \Gamma_{H\rightarrow d \overline{d}s \overline{s}}+ 6\cdot \Gamma_{H\rightarrow u \overline{u}s \overline{s}}\\
\hspace*{2.6cm}+ 2\cdot \Gamma_{H\rightarrow u \overline{u}u \overline{u}}+ 3\cdot \Gamma_{H\rightarrow d \overline{d}d \overline{d}}\ ,
$
\\
\\
$
\Gamma_{WW/ZZ-int} = 3\cdot \Gamma_{H\rightarrow \nu_e \overline{e}e \overline{\nu_e}} - 3\cdot \Gamma_{H\rightarrow \nu_e\overline{\nu_e}\mu \mu^+} 
- 3\cdot \Gamma_{H\rightarrow\nu_e\overline{e} \mu\overline{\nu_{\mu}}} + \\
\hspace*{3.1cm}2\cdot \Gamma_{H\rightarrow u \overline{d}d \overline{u}} -
2\cdot \Gamma_{H\rightarrow u \overline{u}s \overline{s}} - 2\cdot \Gamma_{H\rightarrow u \overline{d}s \overline{c}}\ .
$

	Using scenarios obtained from the experimental measurements, \cite{8.}, we presented the calculated NLO results on the four-fermion decays of light $CP$-even Higgs boson, $h$, Table 2
\vspace{5 mm}
\bec
\emph{\textbf{Table 2.}} {\it  Decay widths of lightest Higgs boson, $h$.
}
\ec
\bec
\begin{tabular}{@{}llllc@{}}\hline
    & \multicolumn{1}{p{1.5cm}}{\centering \tiny\bf{Full decay width of lightest Higgs boson, $h$, (MeV)}} & $\Gamma \rightarrow WW$ & $\Gamma \rightarrow ZZ$ & $\Gamma^{int}$ \\ 
\hline
5-B1 	 & 0.92852      & 0.8326 & 0.1007   &  -0.00478 \\
7-B1     & 0.93026      & 0.8311 & 0.104 & -0.00484   \\  
\end{tabular}
\ec
\vspace{0 mm}

	We also perform calculations of CP-even Higgs boson with the different parameters presented below, in Table 3 and decay widths, Table 4
\vspace{5 mm}
\bec
\emph{\textbf{Table 3.}} {\it  THDM input parameters.
}
\ec
\bec
\begin{tabular}{@{}lllllll@{}}\hline
    & \small\bf $M_H$, GeV & \small\bf $M_{H+}$, GeV & \small\bf $M_A$, GeV &  $\lambda_5$ & \small\bf $\tan \beta$ & $c_{\alpha \beta}$ \\ 
\hline
I 	 & 360    & 690 & 420  & -1.9 & 4.5 & 0.15 \\
II   & 600    & 690 & 690  & -1.9 & 4.5 & 0.15 \\  
\end{tabular}
\ec
\vspace{0 mm}
\vspace{5 mm}
\bec
\emph{\textbf{Table 4.}} {\it  Decay width of $CP$-even Higgs boson, $H$.
}
\ec
\bec
\begin{tabular}{@{}lllc@{}}\hline
 \multicolumn{1}{p{1.5cm}}{\centering \tiny\bf{Full decay width of lightest Higgs boson, $h$, (MeV)}} & $\Gamma \rightarrow WW$ & $\Gamma \rightarrow  ZZ$ & $\Gamma^{int}$ \\ 
\hline
 -54.487      & -71.47 & 17.203   &  -0.22 \\
 1176.36      & 789.98 & 385.74 & 0.64   \\  
\end{tabular}
\ec
\vspace{0 mm}

The calculations of SM Higgs boson decay width give us the following result, Table 5
\vspace{5 mm}
\bec
\emph{\textbf{Table 5.}} {\it  Decay width of SM Higgs boson, $H_{SM}$
}
\ec
\bec
\begin{tabular}{@{}lllc@{}}\hline
 \multicolumn{1}{p{1.5cm}}{\centering \tiny\bf{Full decay width of lightest Higgs boson, $h$, (MeV)}} & $\Gamma \rightarrow WW$ & $\Gamma \rightarrow ZZ$ & $\Gamma^{int}$ \\ 
\hline
 0.958     & 0.858 & 0.10724   &  -0.00724 \\
\end{tabular}
\ec
\vspace{0 mm}

In the absence of beyond SM (BSM) Higgs decay modes, total scale factor $\kappa_H^2$ is the value of the deviation of the sum of the partial Higgs decay widths compared to the SM total width $\Gamma^{SM}_H$, \cite{9., 10.}:

\[
\kappa_H^2 \left( \kappa_i,m_H \right) = \sum\limits_{{j=WW^{\star}, ZZ^{\star}, b\overline{b}, \tau^-\tau^+,\gamma\gamma, Z\gamma, gg, t\overline{t},c\overline{c}, s\overline{s},  \mu^-\mu^+}} \frac{\Gamma_j( \kappa_i,m_H)}{\Gamma^{SM}_H( m_H)} \ .
\]

Since the identification of four leptons is the most detectable decay mode in comparison with other decay channels, the optimal direction of the search for new physics will be finding and comparison of the factor $\kappa_H^2$ for the decays of two Higgs bosons --- the lightest and the heaviest one into $WW$ or $ZZ$ bosons.
So, the scale factor $\kappa_H^2$  in this case is the following:
\[
\kappa_H^2 \left( \kappa_i,m_H \right) = \sum\limits_{{j=WW^{\star}, \ ZZ^{\star}}} \frac{\Gamma_j( \kappa_i,m_H)}{\Gamma^{SM}_H( m_H)}\ .
\]
It is also interesting to calculate scale factors $\kappa_W^2$ and $\kappa_Z^2$

\[
\frac{\Gamma_{WW^*}}{\Gamma^{SM}_{WW^*}} = \kappa_W^2,
\]
\[
\frac{\Gamma_{ZZ^*}}{\Gamma^{SM}_{ZZ^*}} = \kappa_Z^2,
\]
which allow probing for BSM contributions in the loops for each channel separately. Moreover, these factors make it possible to calculate the deviations from the SM of the associated production cross sections in accordance with the formulas:	
\[
\frac{\sigma_{WH}}{\sigma^{SM}_{WH}} = \kappa_W^2,
\]
\[
\frac{\sigma_{ZH}}{\sigma^{SM}_{ZH}} = \kappa_Z^2.
\]
The results of our calculations are performed in the Table 6:

\vspace{7 mm}
\bec
\emph{\textbf{Table 6.}} {\it  Scaling factors  of two Higgs bosons 
}
\ec
\bec
\begin{tabular}{@{}lcllll@{}}\hline
 \multicolumn{1}{}{\small\bf{Higgs boson}} & \small\bf {Scenario} & $\kappa_W^2$ & $\kappa_Z^2$ & \multicolumn{1}{p{1.2cm}}{\centering $\kappa_H^2$ \\ w/o int} & \multicolumn{1}{p{1cm}}{\centering $\kappa_H^2$ \\ w int} \\ 
\hline
h & 5-B1 & 0.97 & 0.939 & 0.967 & 0.969 \\
h & 7-B1 & 0.968 & 0.97 & 0.969 & 0.971 \\
H & II & 921 & 3597 & 1218 & 1228 \\
\end{tabular}
\ec
\vspace{7 mm}
		
	From the comparison of the data from Table 6 we see the slight change in $\kappa_H^2$ factor for 5-B1 and 7-B1 scenarios and huge increase compared to SM  one for scenario II. Moreover, we can see the increasing of $\kappa_H^2$ factor for all scenarios with inclusion of interference. The BSM contributions in the loops for $WW$ channel are larger in 5-B1 scenario but for 7-B1 scenario the larger contribution in the loops are for $ZZ$ channel. Therefore, the chose of renormalization schema is also essential to the final result. The sharp jump in the $\kappa_H^2$ factor for heavier $CP$-even Higgs boson indicates about significant deviation from the $SM$ for scenario 7-B1. The difference in factor $\kappa_Z^2$ compared to $\kappa_W^2$ by almost four times indicates the predominance of the associated with $Z$ boson production cross section of $CP$-even Higgs boson over the associated with $W$ production cross section.

\begin{center}
\textbf{\textsc{3. Conclusions}}
\end{center}

The searches for BSM physics are connected with studying of Higgs boson properties. The way of the realization of this purpose is connected with the decay widths measurements and theoretical predictions of Higgs boson properties. The most perspective and convenient Higgs boson decay channel into four fermions is one of the interesting way of its investigation. For the precise measurements of the decay width is proposed THDM model in the paper. We have considered lightest and CP-even heavier Higgs bosons, h and H correspondingly and modeled their decay widths into four fermions with the help of Monte Carlo program Prophecy 4f 3.0. The results of our calculations led us to the following conclusions connected with the searches of deviations from SM:

\begin{itemize}
\item decay widths of lightest Higgs boson, $h$ and $H_{SM}$ almost do not differ from each other;
\item the scale factor $\kappa_H^2$ of $CP$-even Higgs boson, $H$ equal to 1228;
\item the calculations of decay widths strongly depend on the parameter space and can take negative values as the masses of the $CP$-even and $CP$-odd Higgs bosons decrease by almost two times from the parameters of the 7B1 scenario;
\item   the interference account leads to an insignificant increase in decay widths;
\item the difference in factor $\kappa_Z^2$ compared to $\kappa_W^2$ by almost four times indicates the predominance of the associated with $Z$ boson production cross section of $CP$-even Higgs boson, $H$ over the associated with $W$ production cross section.
\item BSM contributions in the loops for $WW$ and $ZZ$ channels are vary depending on renormalization schema.
\end{itemize}

\end{document}